\documentclass[twocolumn,prl,superscriptaddress,showpacs,10pt,aps]{revtex4-1}

\usepackage{amsmath}
\usepackage{amssymb}
\usepackage{graphicx}
\usepackage{color}

\usepackage[T1]{fontenc}
\usepackage[utf8]{inputenc}

\usepackage{amsthm}
\usepackage{bbm}
\usepackage{mathtools}

\def\tr{\operatorname{tr}}
\def\idty{{\leavevmode\rm 1\mkern -5.4mu I}} %  unit operator

\def\Rl{{\mathbb R}}

\def\ketbra #1#2{{\vert#1\rangle\langle#2\vert}}
\def\kettbra#1{\ketbra{#1}{#1}}

\def\tr{\mathop{\rm tr}\nolimits}
\def\abs#1{\vert#1\vert}

\let\veps\varepsilon

\begin{document}
\title{Proof of Heisenberg's error-disturbance relation}

\author{Paul Busch}
\email{paul.busch@york.ac.uk}
\affiliation{Department of Mathematics, University of York, York, United Kingdom}

\author{Pekka Lahti}
\email{pekka.lahti@utu.fi}
\affiliation{Turku Centre for Quantum Physics, Department of Physics and Astronomy, University of Turku, FI-20014 Turku, Finland}

\author{Reinhard F. Werner}
\email{reinhard.werner@itp.uni-hannover.de}
\affiliation{Institut f\"ur Theoretische Physik, Leibniz Universit\"at, Hannover, Germany}

\date{\today}
\begin{abstract}
While the slogan ``no measurement without disturbance'' has established itself under the name {\em Heisenberg effect} in the
consciousness of the scientifically interested public, a precise statement of this fundamental feature of the quantum world has 
remained elusive, and serious attempts at rigorous formulations of it as a consequence of quantum theory have led to seemingly 
conflicting preliminary results. Here we show that despite recent claims to the contrary 
[Rozema {\em et al}, {\em Phys. Rev. Lett.} {\bf 109}, 100404 (2012)],
 %\cite{Roz12},
Heisenberg-type inequalities can be proven that describe a trade-off between the precision of a position
measurement and the necessary resulting disturbance of momentum (and vice versa). More generally, these inequalities are 
instances of an uncertainty relation for the imprecisions of {\em any} joint measurement of position and momentum. Measures of error and disturbance are here defined as figures of merit characteristic of measuring devices. As such they are state independent, 
each giving worst-case estimates across all states, in contrast to previous work that is concerned with the relationship between error and disturbance in an individual state.
\end{abstract}

\pacs{03.65.Ta, % Foundations of quantum mechanics; measurement theory
      03.65.Db, % Functional analytical methods in QM
      03.67.-a 	% Quantum information 
}
\maketitle

%Intro
In spite of their important role since the very beginning of quantum mechanics, uncertainty relations have recently become the subject of active scientific debates. On one hand, entropic versions of the information-disturbance trade-off \cite{entropicUR} have become an important tool in security proofs \cite{furrer} for continuous variable cryptography. On the other hand there were widely publicized \cite{publicity} claims of a refutation \cite{Ozawa04,Erh12,Roz12} of the error-disturbance uncertainty relations heuristically claimed by Heisenberg \cite{Heisenberg1927}. A review of the literature on uncertainty relations is given in \cite{BuHeLa07}.

Heisenberg's 1927 paper \cite{Heisenberg1927} introducing the uncertainty relations is one of the key contributions to early quantum mechanics. It is part of virtually every quantum mechanics course, almost always in the version forwarded by Kennard \cite{kennard}, Weyl \cite{Weyl} and Robertson \cite{robertson}. What is often overlooked, however, is that this popular version is only one way of making the idea of uncertainty precise. The original paper begins with a famous discussion of the resolution of microscopes, in which the accuracy (resolution) of an approximate position measurement is related to the disturbance of the particle's momentum. 

This situation is no way covered by the standard relations, since in an experiment concerning the Kennard-Weyl-Robertson inequality no particle meets with both a position and a momentum measurement. Heisenberg's semiclassical discussion has no immediate translation into the modern quantum formalism, particularly since the momentum disturbance {\em prima facie} involves the comparison of two (generally) non-commuting quantities, the momentum before and after the measurement. Such a translation does require some careful conceptual work, and one can arrive at different results. This is shown by the example of Ozawa \cite{Ozawa04}, who defines a relation he claims to be a rigorous version of Heisenberg's ideas, and shows that it fails to hold in general. A suggested modification of the false relation has recently been verified experimentally \cite{Erh12,Roz12}. This has been widely publicized as a refutation of Heisenberg's ideas, in apparent contradiction to our main result. However, there is no contradiction, and the disagreement only shows that there is a grain of rigorously explicable truth in Heisenberg, provided one looks in the right place for it. While Ozawa aims to describe the interplay between error and disturbance for an individual state,
our approach gives a state-independent characterization of the overall performance of measuring devices.
% error and disturbance. 
In \cite{BLW2013a} we show that Ozawa's notions, though mathematically well-defined, have only limited validity as measures of
error and disturbance.\footnote{This work is a detailed elaboration of criticisms that were raised by the present authors in earlier
publications, for instance, in \cite{BuHeLa04} and \cite{Werner04}.}
%A detailed critique of Ozawa's approach will be given elsewhere .
 
We will describe and prove an inequality of the classic form
\begin{equation}\label{UR}
 (\Delta Q)(\Delta P)\geq\frac\hbar2\ ,
\end{equation}
in which the quantities $\Delta Q$ and $\Delta P$ are {\em not} given by the variances of the position and momentum distributions in the same state, as in the textbook inequality. Instead, following closely the suggestion of Heisenberg, they are explicitly defined figures of merit for a microscope-like measurement scenario: the accuracy $\Delta Q$ of a position measurement and the momentum disturbance $\Delta P$ incurred by it.  Moreover, the inequality is sharp, and we will describe explicitly the cases of equality. We believe that the definitions and results are simple enough to use in a basic quantum mechanics course, although the full proof uses some tools beyond such a course.

The main progress over earlier work \cite{Werner04} is a simpler definition of the $\Delta$ quantities, using the idea of calibration \cite{BuPe07}. This definition does not require the Monge transportation metric, which led in \cite{Werner04} to quantities akin to absolute deviations rather than root mean square deviations, and hence to a constant different from $\hbar/2$ in \eqref{UR}. A changed constant (even if optimal for the particular definitions of $\Delta$) puts an undue burden on the memory of undergraduates. Using variances also for calibration solves this problem. The basic ideas of the proof in \cite{Werner04} can be taken over.

\begin{figure}[ht]
\centering
  \includegraphics[width=3.3cm]{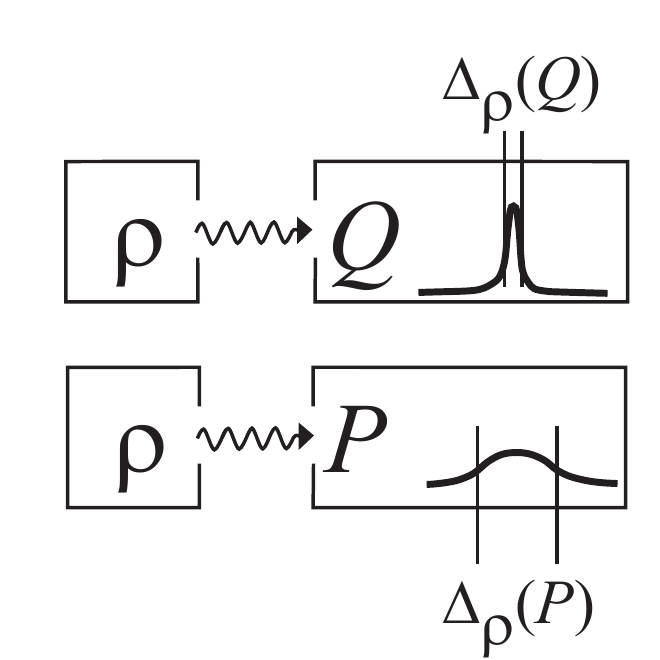}
	\caption{Scenario of preparation uncertainty. $\Delta_\rho$ is the root of the variance of the distribution obtained for the indicated observable in the state $\rho$.
     In this pair of experiments no particle is subject to both a position and a momentum measurement. }
	\label{fig:PUR}
\end{figure}

%\section{Preparation vs. measurement uncertainty}
To keep matters simple, we stick to the classic situation of two canonically conjugate variables of a single quantum degree of freedom. For the sake of comparison, let us recall the scenario of the Kennard-Weyl-Robertson inequality, which we call preparation uncertainty (see Fig.~\ref{fig:PUR}).
The spreads $\Delta_\rho(A)=\bigl(\tr\rho A^2-(\tr\rho A)^2\bigr)^{1/2}$ of position $Q$ and momentum $P$ are determined in separate experiments on the same source, given by a density operator $\rho$. The uncertainty relation $\Delta_\rho(Q)\Delta_\rho(P)\geq\hbar/2$ is a quantitative version of the observation that there are no dispersion-free quantum states \cite{vonNeumann}, as applied to a canonical pair of observables. It is {\it not} to be found in Heisenberg's paper \cite{Heisenberg1927}, except in a rough discussion of post-measurement states, which he assumes to be Gaussian with a spread related to the accuracy of a position measurement.

\begin{figure}[ht]
\centering
  \includegraphics[width=5cm]{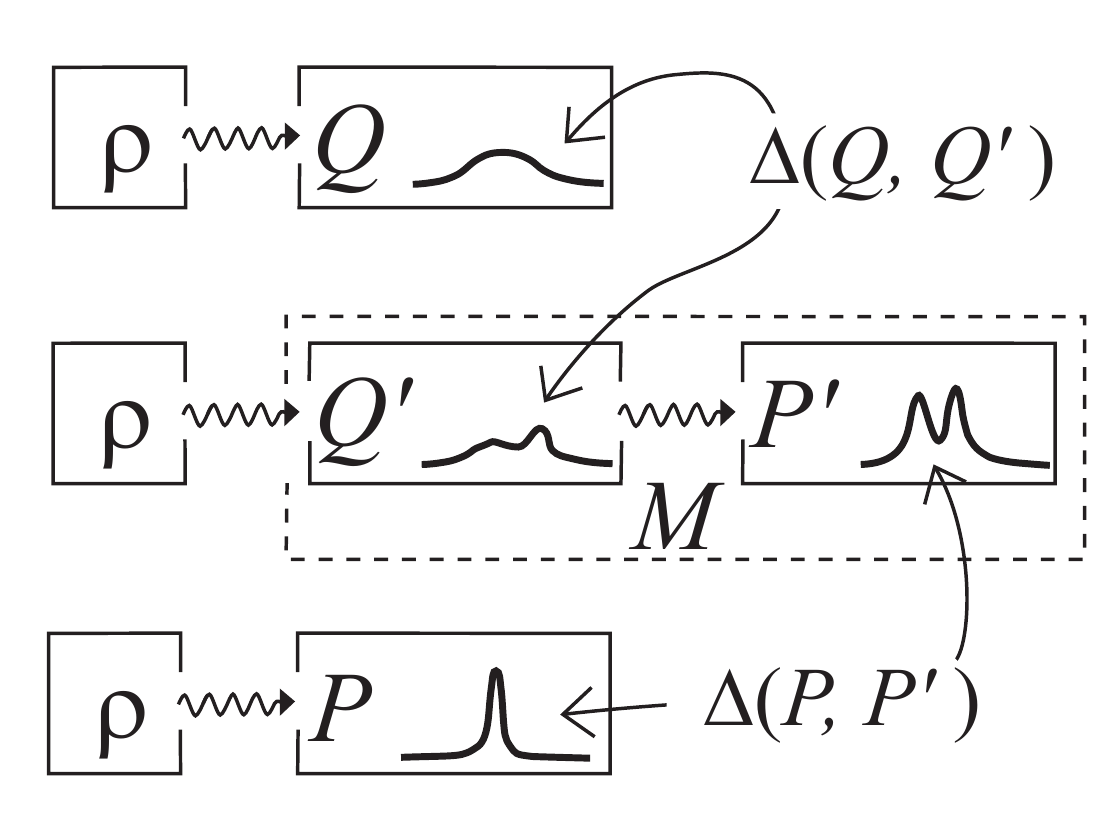}
	\caption{Scenario of measurement uncertainty for successive measurements, as discussed by Heisenberg (middle row). An approximate position measurement $Q'$ is followed by an ideal momentum measurement,
    effectively given a measurement $P'$ on the initial state. The accuracy $\Delta(Q,Q')$ quantifies the difference between the output distributions of $Q'$ and an ideal position measurement $Q$ (first row).
    Similarly, the momentum disturbance $\Delta(P,P')$ quantifies the difference between the distributions obtained by $P'$ and by an ideal momentum measurement $P$ (last row).
    The definitions for these $\Delta$ quantities (see text) can be applied, more generally, to an arbitrary joint measurement $M$ (dashed box). This can be any device producing, in every shot,
    a $q$ value and a $p$ value. $Q'$ and $P'$ are then defined as the marginals of $M$, obtained by ignoring the other output. }
	\label{fig:MUR}
\end{figure}

In contrast, Fig.~\ref{fig:MUR} shows the scenario discussed by Heisenberg. The middle row shows an approximate position measurement $Q'$ followed by a momentum measurement.
How should we  define the momentum disturbance and position error in this setup? The error of the approximate position measurement $Q'$ clearly refers to the comparison with an ideal measurement $Q$ as shown in the first row.  For the momentum disturbance we can say the same: We have remarked that the momenta before and after the microscope interaction do not commute, so the difference makes no sense in the individual case. However, we can compare the {\em distributions} of the momenta measured after the position measurement (call this effective measurement $P'$) with the {\em distribution} an ideal momentum measurement $P$ would have given on the same input state. Come to think of it, this is precisely how we detect disturbance in other typical quantum settings.
Consider, for example, the double slit experiment. It is well-known that illuminating the slits enough to detect the passage of a particle through one or the other hole makes the interference fringes go away.  Clearly, the light used for observation disturbs the particles, and the evidence for this is once again the change of the {\it distribution} on the screen. Note that this way of looking at error and disturbance restores the symmetry between the position and momentum aspects of this scenario. The uncertainty relations we will prove therefore apply just as well to the position disturbance caused by an approximate momentum measurement and, more generally, to any measurement scheme $M$, which produces in every run a value $p$ and a value $q$ (see the dashed outline in Fig.~\ref{fig:MUR}). This generalization also covers any successive measurement scenario, in which one tries to {\em correct for} some of the momentum disturbance, perhaps using the detailed knowledge of how the position measuring device works. In principle, this could allow a reduction of uncertainties. However, the inequality holds without change, which gives a precise meaning and a proof to Heisenberg's phrase ``uncontrollable momentum disturbance'', which he himself uses without further justification.

Let us now discuss the definition of $\Delta(Q,Q')$ in more detail (the momentum case will be completely analogous). We think of this ``microscope resolution'' as a figure of merit for the device, a promise which might be advertised by the manufacturer, and which could be verified by a testing lab. $\Delta(Q,Q')=0$ will mean that the ``approximate'' device $Q'$ is completely equivalent to the ideal $Q$, i.e., for every input state $\rho$ the output distributions will be the same. Similarly, a small value might indicate that the difference in the distributions will be small for every input state. This requires a definition for the distance of two general probability distributions, which we will give below (Section on ``Uncertainty metrics''). However, we can also take a simpler approach, which avoids verifying a statement for {\it all} input states. Instead the testing lab might concentrate on those states, which at least classically would seem to be the most demanding ones, namely states for which $Q$ has a known and sharp value. We call this process ``calibration''. Still, this requires testing of many states
but no longer on very mixed states, or states which contain coherent superpositions of widely separated wave functions.

An advantage of the calibrated error is that we no longer need a quantitative evaluation of the distance between arbitrary probability distributions, but just between an arbitrary distribution and a known sharp value $\xi$. For this we naturally take the root mean square deviation from $\xi$
\begin{equation}\label{qdev}
  D(\rho,Q';\xi)=\Bigl\langle(q'-\xi)^2\Bigr\rangle_{\rho,Q'}^{1/2}
\end{equation}
where the angle brackets denote the expectation of the indicated function of the output $q'$, in the distribution obtained on the preparation $\rho$ with the device $Q'$. This statement allows for $Q'$ to be a general positive operator valued measurement.  For projection valued observables like $Q$ we could simplify this to $D(\rho,Q;\xi)^2=\tr\bigr(\rho(Q-\xi\idty)^2\bigl)$. The latter quantity is to be small, say $\leq\veps$, for the input states $\rho$ used for calibration. Hence we set $\Delta_c(Q,Q')$ to be  %\cite{note2}
\begin{equation}\label{Deltac}
 \lim_{\veps\to0}\,\sup\Bigl\{D(\rho,Q';\xi)\Bigm|\rho,\xi;\  D(\rho,Q;\xi)\leq\veps \Bigr\}.
\end{equation}
Here the set is non-empty since for any $\xi$ and $\epsilon >0$ there is a $\rho$ such that  $\tr(\rho\, Q)=\xi$ and $D(\rho,Q;\xi)<\epsilon$; moreover,
the limit exists, because with decreasing $\veps$ the supremum is over fewer and fewer states, so the function is non-increasing.
In the case of a bad approximation, the supremum can be infinite, in which case we put $\Delta_c(Q,Q')=\infty$.

With this definition, and the corresponding one for $P$, we can state our main result. We just assume that the $Q'$ and $P'$ are the marginal observables of some joint measurement device $M$ whose calibration errors are both finite. As discussed above this covers also the case of a sequential measurement (Fig.~\ref{fig:MUR}). Then
\begin{equation}\label{MUR}
  \Delta_c(Q,Q')\, \Delta_c(P,P')\geq\frac\hbar2\ .
\end{equation}
This inequality is sharp, and equality holds for an $M$ for which the joint distribution of $(q,p)$-outputs is the so-called Husimi distribution \cite{Husimi} of the input state, which can be obtained by a Gaussian smearing of the Wigner function. In the extreme case of one of the marginals being error free, the error for the other marginal is necessarily infinite.

%\subsection*
\noindent {\em Proof.\ }
The proof has two parts: The first is elementary and concerns the special case that $M$ is a {\it covariant phase space observable}. These observables \cite{Davies,Holevo,QHA,Husimi} can be described explicitly, including a very simple form of their marginals $Q'$ and $P'$, by which \eqref{MUR} can be reduced to the preparation uncertainty. The second, more technical part of the proof reduces the general case to the covariant case by an averaging method, and is taken from \cite{Werner04}. We only sketch it \footnote{A full proof is given in \cite{BLW2013b}. This paper also generalizes the preparation and measurement uncertainty relations from root-mean-square deviations to power-$p$ deviations.}.

By a covariant measurement we mean one which has a natural symmetry property for both position and momentum translations. That is, if we apply it to an input state shifted in position by $\delta q$ and in momentum by $\delta p$, the output distribution will be the same as before, transformed by $(q,p)\mapsto(q+\delta q,p+\delta p)$. These symmetries are implemented by the Weyl operators (a.k.a. Glauber translations) $W(q,p)=\exp((iqP-ipQ)/\hbar)$. Then the whole observable can be reconstructed from its density at the origin, which must be \cite{Holevo,QHA} a positive operator $\sigma$ of trace $1$, i.e., a density operator as for a quantum state. The probability for outcomes in a set $S\subseteq\Rl^2$ is then given by the positive operator
\begin{equation}\label{povm}
  M(S)=\int_S\mskip-5mu \frac{dq\,dp}{2\pi\hbar}\ %\chi_S(q,p\ )
  W(q,p)^*\sigma W(q,p) \ . %\ ,
\end{equation}
 %where $\chi_S$ denotes the indicator function, which is $1$ for $(q,p)\in S$ and $0$ otherwise.
 A remarkable property of these joint measurements of position and momentum is that their marginals take a particularly simple form: The probability density of the outputs $q'$ obtained on a state $\rho$ is a convolution of the position distributions of $\rho$ and $\sigma$. That is, we can model the output distribution by taking $q$ distributed like the outputs of an ideal measurement $Q$ on $\rho$, and adding a noise term $q''$, which is independent of $q$ and distributed according to the position distribution of $\sigma$. The same description applies to the marginal $P'$.

Therefore, for a covariant measurement we can immediately identify $\Delta_c(Q,Q')$ without further computation: The density $\sigma$ is a fixed characteristic property  of the measurement. Therefore, as the position distribution of $\rho$ becomes sharply concentrated around some $\xi$, the outputs converge in distribution to $q'=\xi+q''$, so
\begin{equation}\label{DelCov}
  \Delta_c(Q,Q')=D(\sigma,Q;0) \ ,
\end{equation}
which is the ``size'' (the root mean square deviation) of the ``noise''. For example, if $\sigma$ has sharp position distribution at some value $a$, this is equal to $|a|$, since the outputs will be off by a shift $a$  (i.e., $q'\approx q+a$). Hence one will choose $\sigma$ with zero mean. The uncertainty product then becomes
$\Delta_c(Q,Q')\Delta_c(P,P')=\Delta_\sigma(Q)\Delta_\sigma(P)$, which is $\geq\hbar/2$ by the preparation uncertainty relation applied to $\sigma$. This proves \eqref{MUR} for the case of covariant measurements, and at the same time provides examples of minimum uncertainty measurements: all we have to do is to choose $\sigma$ as a centered minimum uncertainty state, i.e., as $\sigma=\kettbra\Psi$ with $\Psi$ a real valued centered Gaussian wave function. The phase space distribution associated with an input state $\rho$ by this measurement $M$ is then the Husimi distribution \cite{Husimi}.

The more technical part of the proof of \eqref{MUR} is to show that for any measurement $M$ there is a covariant one, say $\overline M$ with at most the same $\Delta$'s. Basically, $\overline M$ is obtained from $M$ by averaging, the technical problem being that the parameter range of $(q,p)$ over which one has to ``average'' is infinite (see \cite{Werner04}). Let us introduce
$\mathcal M_{\veps}(\Delta Q,\Delta P)$ as the set of measurements $M$ such that, for $A=Q,P $, $D(\rho,A';\xi)\leq\Delta A$ whenever $D(\rho,A;\xi)\leq\veps$ for given $\Delta A$ and $\veps$. This is a convex set, and compact in a suitable weak topology. We can write the covariance condition as a fixed point equation for some transformations on the set of all observables, namely a unitary transformation by a Weyl operator combined with a shift in the argument. These transformations commute, and leave $\mathcal M_{\veps}(\Delta Q,\Delta P)$ invariant. Therefore, by the Markov-Kakutani fixed point theorem this set, if non-empty, must also contain a covariant element, which by construction has at most the same uncertainties.  This concludes our sketch of the proof of \eqref{MUR}.

%\subsection*
\noindent {\em Uncertainty metrics.\ }
The calibration criterion only involves highly concentrated states so that, in principle, on general input states the optimal joint measurement might produce output distributions quite different from the ideal ones. One can easily give examples of a projection valued observable $A$ and an ``approximation'' $A'$ for which the calibrated distance is a rather optimistic estimate. That is if we denote by $\Delta(Q,Q')$ a figure of merit based on comparison of {\em all} states we might have  $\Delta(Q,Q')\gg\Delta_c(Q,Q')$. Note first that in the covariant case this cannot happen: The statement that $Q'$ can be simulated by adding fixed independent noise to $Q$ is valid for arbitrary input states, and any reasonable definition of $\Delta(Q,Q')$ should give the size of the noise. However, in the general case we would need a definition which is independent of that special form. Here we will introduce such
a quantity and show that an uncertainty holds for it.

The idea is to define a metric $D$ on probability distributions which extends \eqref{qdev} in the sense that $D(\rho,Q';\xi)$ becomes the metric distance between the output distribution of $Q'$ and a point measure at $\xi$. Then we set
\begin{equation}\label{DelAll}
  \Delta(Q,Q')=\sup_\rho D(\rho,Q;\rho,Q'),
\end{equation}
where the expression on the right is the metric distance of the two output distributions. Since $\Delta_c$ takes the supremum over the smaller set of highly concentrated states, we have $\Delta(Q,Q')\geq\Delta_c(Q,Q')$. The metric $D$ on probability distributions is basically fixed by our requirements as what is technically known as the Wasserstein-2 distance, which is a variant of the the Monge-Kantorovich transport or ``earth mover's'' distance (see \cite{Villani} for a study of such metrics). The problem addressed by Monge was the cost of transforming a hill (earth distribution $\mu$) into some fortifications (earth distribution $\eta$), when the workers had to be paid by the bucket and the distance covered. A transport plan, also known as a {\em coupling} between the measures $\mu$ and $\eta$ would be a measure $\gamma$ on $\Rl\times\Rl$ describing how much earth was to be moved from $x$ to $y$. This entails that the marginals of $\gamma$ must be $\mu$ and $\eta$. The cost in the Monge problem is $\int\gamma(dx\,dy)\abs{x-y}$, which is then minimized by choosing an optimal $\gamma$. In the Wasserstein-2 distance the cost function is chosen to be quadratic in the distance and an overall root is taken to bring the units back to a length:
\begin{equation}\label{wass2}
  D(\mu;\eta)=\inf_\gamma\Bigl(\int\gamma(dx\,dy)\abs{x-y}^2\Bigr)^{1/2},
\end{equation}
where the infimum is over all couplings $\gamma$. Consider now the case that $\eta$ arises from $\mu$ by adding independent noise with distribution $\nu$, which amounts  to the convolution $\eta=\mu\ast\nu$. This immediately suggests a transport plan, namely shifting each individual element of the $\mu$ distribution by the amount suggested by the noise (formally: $\gamma(dx\,dy)=\mu(dx)\nu(d(y-x))$). This may not be optimal, but gives the estimate
$D(\mu;\mu\ast\nu)\leq D(\nu;0)$, the size of the noise, where once again the second argument stands for the point measure at zero. This says that the largest distance is attained for a point measure $\mu$, and therefore
\begin{equation}\label{DDc}
  \Delta(Q,Q')=\Delta_c(Q,Q')
\end{equation}
whenever $Q'$ is the marginal of a covariant measurement. To summarize this section: if we define the deviation between $Q$ and $Q'$ by a worst case figure of merit over {\em all} states, the uncertainty relation once again holds. Moreover, the two notions coincide on all covariant measurements, and in particular for the cases of equality.

%\subsection*
\noindent {\em Conclusion and Outlook.\ }
With the inequality \eqref{MUR} we have provided a general, quantitative quantum version of Heisenberg's original semiclassical uncertainty discussion. This is a remarkable vindication of Heisenberg's intuitions, far beyond the usual view, which takes the quantitative content of the paper to be summarized entirely by the preparation inequality, and sees the discussion of the microscope as no more than a heuristic order of magnitude argument.

Our conceptual framework applies to any pair of observables which are not jointly measurable. However, evaluating the respective uncertainty bounds, which will typically not be expressed in terms of the product of uncertainties, is another matter 
requiring further studies.
%We have already expressed our dissatisfaction with the proposed formalization of Heisenberg's ideas by  \cite{Ozawa04}. A more %detailed analysis is in preparation \cite{BuLaWe13}.

%\subsection*
\noindent{\em Acknowledgements.}
Part of this work (P.L.) is supported by the Academy of Finland, project no 138135. R.F.W.\ acknowledges support from the European network SIQS. P.B.\ has been supported by COST Action MP1006.
We wish to thank two anonymous referees for their valuable criticism and recommendations.

\bibliography{UR_Letter}

\end{document}